\documentclass[conference]{IEEEtran}

\usepackage{color}
\usepackage{subfigure}
\usepackage{graphicx}
\usepackage{amsmath}
\usepackage{wrapfig}
\usepackage{algorithmic, algorithm}
\usepackage{multirow}

\begin{document}

\title{Online multipath convolutional coding for real-time transmission}

\author{
\IEEEauthorblockN{Tuan Tran Thai, Emmanuel Lochin, J\'er\^ome Lacan}
\IEEEauthorblockA{University of Toulouse; T\'eSA; ISAE/DMIA; Toulouse, France}
}

\maketitle

\begin{abstract}
Most of multipath multimedia streaming proposals use Forward Error Correction (FEC) approach to protect from packet losses. However, FEC does not sustain well burst of losses even when packets from a given FEC block are spread over multiple paths. In this article, we propose an online multipath convolutional coding for real-time multipath streaming based on an on-the-fly coding scheme called Tetrys. We evaluate the benefits brought out by this coding scheme inside an existing FEC multipath load splitting proposal known as Encoded Multipath Streaming (EMS). We demonstrate that Tetrys consistently outperforms FEC in both uniform and burst losses with EMS scheme. We also propose a modification of the standard EMS algorithm that greatly improves the performance in terms of packet recovery. Finally, we analyze different spreading policies of the Tetrys redundancy traffic between available paths and observe that the longer propagation delay path should be preferably used to carry repair packets.

\end{abstract}

\IEEEpeerreviewmaketitle

\section{Introduction}

Multipath streaming has gained much attention recently thanks to overlay networks and multiple access technologies (e.g., Wi-Fi, Cellular) available by default in handheld devices. The benefits of multipath overlay routing and multipath streaming are presented in \cite{bestpath_vs_multipath}, \cite{Golubchik02multipath} (e.g., reduction in correlation between consecutive packet losses, throughput gain, ability to react to congestion variation in different parts of the network). Another interesting property of multipath has been illustrated in \cite{Fashandi}. Fashandi et al. \cite{Fashandi} showed that the loss rate after packet recovery decays exponentially with the number of paths. However, the challenging task in multipath streaming is to split the data flow among available paths to achieve better perceived video quality. As a potential solution, in \cite{Jurca_mediaflow}, Jurca et al. proposed a load splitting scheme based on an end-to-end (E2E) distortion model for single layer video streaming. Later in \cite{Jurca_forwarderror}, they proposed a similar E2E distortion model for scalable video streaming as an objective function and used optimization algorithms to minimize the distortion. One of the most achieved algorithms is Encoded Multipath Streaming (EMS) framework proposed by Chow et al. \cite{EMS09}. In their proposal, the receiver observes the loss rate on each path, calculates the overall loss rate after packet recovery and sends the load splitting vector to the sender. However, all these proposals (\cite{Fashandi}-\cite{Nguyen03pathdiversity}) use Forward Error Correction (FEC) to protect the video from losses. The main problem is this block code scheme requires to dynamically adapt its initial parameters and as a result, complex probing and network feedback analysis. Recently, a novel erasure coding approach that prevents such complex configuration has been proposed \cite{arq_nc, Lacan, Tetrys}.

In this paper, we propose to use an on-the-fly erasure coding scheme called Tetrys \cite{Tetrys} to real-time multipath streaming and in particular, inside the EMS framework \cite{EMS09}. The rationale of using this framework is because EMS obtains better performance in terms of computation compared to \cite{Jurca_mediaflow}, \cite{Jurca_forwarderror} \cite{RFS}. We show that enabling Tetrys instead of FEC inside EMS greatly improves the overall performance in terms of packet delivery ratio in both uniform and burst losses. We also study the decoupling between load allocation and redundancy traffic with Tetrys and propose several other measurements with different propagation delay not tackled in \cite{EMS09}. The results show that sending Tetrys repair packets on paths with longer propagation delay increases the packet delivery ratio before the E2E delay constraint in real-time transmission limited to hundreds milliseconds. Furthermore, we improve the EMS scheme to better follow the network dynamics and to reduce the loss rate after packet recovery.

The rest of this article is organized as follows. Section \ref{sec:ems_principle} introduces briefly the EMS scheme. Section \ref{sec:tetrys_principle} presents the basic principle of Tetrys and the decoupling between load allocation and Tetrys redundancy traffic. Section \ref{sec:simulations_and_results} shows the results obtained from Tetrys compared to FEC with different settings and the benefits of decoupling between load allocation and Tetrys redundancy traffic. Section \ref{sec:ems_algorithm_improvement} presents the modified EMS algorithm and results. We conclude and provide future work in section \ref{sec:conclusion}.

\section{EMS principle}
\label{sec:ems_principle}
\begin{figure*}[!htb]
\begin{center}
\includegraphics[scale=0.3]{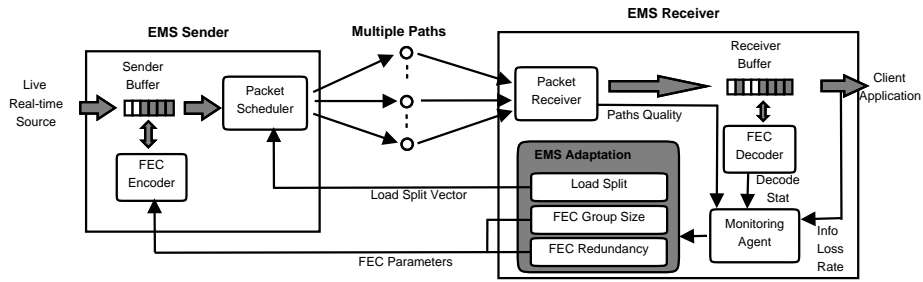}
\caption{EMS overview \cite{EMS09}}
\label{fig:ems_overview}
\end{center}
\end{figure*}

Fig. \ref{fig:ems_overview} shows an overview of the EMS scheme. The FEC encoder in EMS sender takes the live stream and encodes with FEC parameters specified by \textit{k} source packets and \textit{n-k} repair packets. The encoded stream is then splitted among available paths with different characteristics (e.g., propagation delay, loss rate, available bandwidth) thanks to the packet scheduler. The EMS receiver stores all received packets and checks whether it can decode all lost packets in a FEC block specified by FEC(k,n). In the context of live streaming, any packets arrived or recovered after the deadline are discarded.

The EMS scheme is detailed in \cite{EMS09}. Thus, we introduce the most important part of EMS, the Online Load Splitting (OLS). At bootstrap, EMS sender equally splits the load between available paths so that the receiver measures the loss rate in each path. At each period defined by OLS Adapt Window (in second), the EMS receiver executes the OLS algorithm as depicted in the pseudo code \ref{alg:ols_original}. The information loss rate indicates the percentage of data that can not be recovered. After performing the OLS, EMS receiver sends a feedback containing the load splitting vector and the FEC parameters. The packet scheduler of EMS sender follows the load vector upon reception of the feedback.

\begin{algorithm}
\caption{Online Load Splitting (OLS)}
\label{alg:ols_original}
\begin{algorithmic}[1]
\STATE Compute the asymptotic optimal solution and split the load accordingly 
\STATE Sort the paths in the increasing order of loss rate
\REPEAT 
\STATE Pick the first path in the list

\REPEAT 
\STATE Increase the load on the chosen path by pre-defined $\Delta_L$ (3\% by default)
\STATE Decrease the load on each of remaining paths by a fraction of $\delta$, proportional to their respective loss rates
\UNTIL{measured information loss rate increases}
\STATE Remove the chosen path from the list
\STATE Revert to the previous load splitting
\UNTIL{the path list is empty}
\STATE \textbf{goto} Step 2
\end{algorithmic}
\end{algorithm}

It is noted that the load splitting vector only decides the amount of traffic that each path should carry. This inspires our study of decoupling between the load splitting vector and redundancy traffic with Tetrys (will be described in Section \ref{sec:tetrys_decoupling}).

\section{Tetrys multipath} 
\label{sec:tetrys_principle}
We introduce in this section an on-the-fly erasure coding scheme called Tetrys coupled with EMS scheme for real-time multipath streaming. Then, we present the rationale of decoupling between load allocation and Tetrys redundancy traffic.

\subsection{Basic principle of Tetrys}
Tetrys uses an elastic encoding window buffer $B_{EW}$ which includes all the source packets sent without acknowledgment. For every $k$ source packets, Tetrys sender sends a repair packet $R_{(i..j)}$ which is built as a linear combination of all packets currently in $B_{EW}$ from packet indexed \textit{i} to \textit{j}. The receiver is expected to periodically acknowledge the received or decoded packets. Upon reception of acknowledgment, the sender removes the acknowledged packets out of its $B_{EW}$. Generally, the receiver can decode lost packets as soon as the number of repair packets received is equal to the number of lost packets. By this principle, Tetrys is tolerant to burst losses in neither source, repair nor acknowledgment packets as long as the redundancy ratio exceeds the packet loss rate (PLR). Furthermore, the lost packets are recovered within a delay that does not depend on the Round Trip Time (RTT). This property is very important for real-time applications. 

\begin{figure}[!htb]
\begin{center}
\includegraphics[scale=0.40]{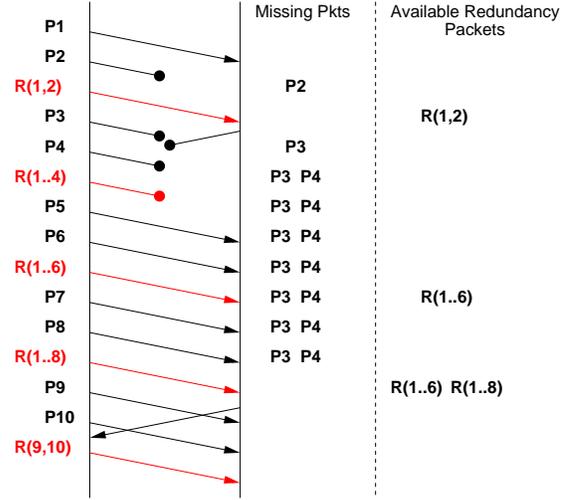}
\caption{A simple data exchange with Tetrys (k=2) \cite{Tetrys}}
\label{fig:tetrys_example}
\end{center}
\end{figure}

Let us show in Fig. \ref{fig:tetrys_example} a simple Tetrys data exchange with $k$ = 2 which implies that a repair packet is sent for every two sent source packets (or redundancy ratio of 33.3\%). The packet $P_2$ is lost during the data exchange. However, the reception of repair packet $R_{(1,2)}$ allows to rebuild $P_2$. The acknowledgment for packets $P_1$ and $P_2$ from the receiver is lost. This loss does not interrupt the transmission, the sender simply continues to compute the repair packets from $P_1$. Later, the lost packets $P_3$, $P_4$ are rebuilt thanks to $R_{(1..6)}$ and $R_{(1..8)}$. The reception of second acknowledgment packet allows the sender to remove the acknowledged packets and build the repair packets from $P_9$. The reader is referred to \cite{Tetrys} for further details. 

\subsection{Decoupling load allocation and redundancy traffic with Tetrys}
\label{sec:tetrys_decoupling}


In \cite{Kurant11}, Kurant showed that the propagation time differences between paths reach several tens of milliseconds by measurements. In case of FEC, the last packet (source or repair) in a FEC block arrived at the receiver must be sooner than the end-to-end (E2E) delay requirements normally specified by the path with longest propagation delay. If the arrival date of last FEC packet exceeds the deadline due to long block size or queuing delay in the network, the sender should reduce the block size. The size reduction makes FEC less tolerant to burst losses (see later in Fig. \ref{fig:fec_compa_gb2}). Thus, we believe that FEC repair packets can be sent to any available paths without changing the result with well dimensioning block size. On the other hand, the arrival time of Tetrys repair packets is rather important since they are used to recover all previous lost packets if possible. If a repair packet built from sent source packets without acknowledgment is transmitted to the path with short propagation delay, it is likely that the repair packet arrives sooner than the source packets sent on longer paths. This means that the arrival of repair packet can not be used to recovered the previous lost packets at its arrival even though the source packets sent on longer paths arrive successfully. This reduces the effectiveness of Tetrys repair packets in real-time streaming. Based on this observation and the independence between load splitting vector and packet scheduler (see section \ref{sec:ems_principle}), we propose to decouple the load allocation on each path specified by the load splitting scheme and the way Tetrys repair packets are sent. This implies that Tetrys repair packets are preferably sent to the path with longer propagation delay while keeping the same load allocation. 

Table \ref{tab:tetrys_strategies} show different strategies to send Tetrys repair packets. For instance, with ``Tetrys long'' strategy, the Tetrys repair packets are first sent to the path with longest propagation delay. If the load on longest path is fulfilled, the Tetrys repair packets are sent to the path with second longest propagation delay and so on. While Tetrys repair packets are sent to the available path according to the packet scheduler in ''Tetrys'' strategy.

\begin{table}[h!]
\caption{Different strategies of sending Tetrys repair packets}
\begin{center}
	\begin{tabular}{|l|l|}
	\hline
		Tetrys long & preferably sent to path with longer delay \\ \hline
		Tetrys short & preferably sent to path with shorter delay \\ \hline
		Tetrys & sent to available path \\ \hline
	\end{tabular}
\end{center}
\label{tab:tetrys_strategies}
\end{table}


\section{Simulations and results}
\label{sec:simulations_and_results}
We use \textit{ns-2} \cite{ns} to evaluate Tetrys and FEC using EMS scheme. The number of paths is specified in each simulation.  These paths can be built thanks to multiple physical interfaces or overlay network. The path establishment is out of scope of this article. We assume that the available bandwidth exceeds the application rate. The one-way E2E delay constraint is set to 150ms based on ITU-T/G.114 \cite{itu} which is recommended for highly interactive applications. One of the main characteristics of Tetrys is to be fully reliable whatever the burst size \cite{Tetrys}. Indeed, all lost packets are recovered if the redundancy ratio exceeds the PLR. However, we consider the packets as lost at the application level if their arrival or recovery date exceeds the deadline. The information loss rate indicates the percentage of lost data that can not be recovered or be recovered after the deadline of 150ms. To simulate the burst losses, we use a Gilbert-Elliot model in \cite{Tetrys} which is specified by an average PLR and an average length of consecutive lost packets (or shortly mean burst size). In each simulation, the streaming server sends a Constant Bit Rate (CBR) traffic at 1900 kb/s with packet size of 210 bytes. The frequency of Tetrys acknowledgment packet is set to 10ms. The feedback frequency does not change the result since it only affects the buffer sizes.


\subsection{Comparison between FEC and Tetrys with the same EMS scheme}
\label{sec:compa_fec_tetrys}
\begin{figure}[!htb]
\begin{center}
\includegraphics[scale=0.70]{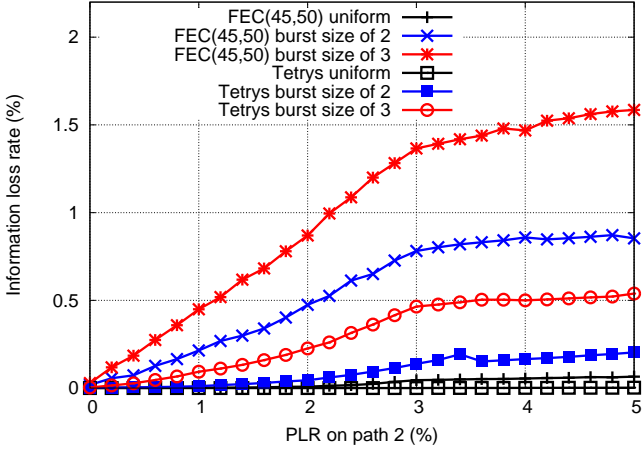}
\caption{Tetrys vs FEC(45,50) at PLR of 3\% on path 1 with uniform losses and burst losses with mean size of 2 and 3 packets, redundancy ratio is 10\%}
\label{fig:tetrys_vs_fec_ems}
\end{center}
\end{figure}

In this simulation, there are two paths between a sender and a receiver. The propagation delay on each path is set to 50ms. The streaming lasts 4 hours and the OLS Adapt Window is set to 60s. The redundancy ratio is set to 10\% which is equivalent to FEC(45,50). The PLR on path 1 is set to 3\% and the PLR on path 2 varies from 0\% to 5\%. Fig. \ref{fig:tetrys_vs_fec_ems} shows that Tetrys consistently outperforms FEC(45,50) in both uniform and mean burst size of 2 and 3 packets. More specifically, Tetrys reduces up to more than 1\% information loss rate in case of mean burst size of 3 packets. With the video coding standard H.264/AVC, the Peak Signal to Noise Ratio (PSNR) with Tetrys can gain up to several dBs \cite{Wenger03}. It is noted that the result with FEC(45,50) in Fig. \ref{fig:tetrys_vs_fec_ems} is similar to Fig. 14 in \cite{EMS09}. In fact, when the PLR on path 2 is less than 3\%, the EMS scheme tends to assign more load on path 2, thus the information loss rate proportionally increases with the PLR on path 2. When the PLR on path 2 is greater than 3\%, the EMS scheme switches to assign more load on path 1. This results in a rather flat in information loss rate at PLR on path 2 greater than 3\%.

EMS scheme comes with FEC redundancy and FEC block size adaptations (see Fig. \ref{fig:ems_overview}). On the other hand, the configuration with Tetrys is simpler than FEC since Tetrys does not need to scale the block size. With the same redundancy ratio, Tetrys achieves smaller information loss rate. Thus, with the redundancy adaptation so that the loss requirement less than a threshold (normally 1\% for video), Tetrys requires less redundancy than FEC. In fact, in Fig. \ref{fig:tetrys_vs_fec_ems}, the information loss rate of Tetrys is much less than 1\% at mean burst size of 3 packets at redundancy ratio of 10\% while the redundancy for FEC must be greater than 10\% to lower its information loss rate to less than 1\%.

\begin{figure}[!htb]
\begin{center}
\includegraphics[scale=0.70]{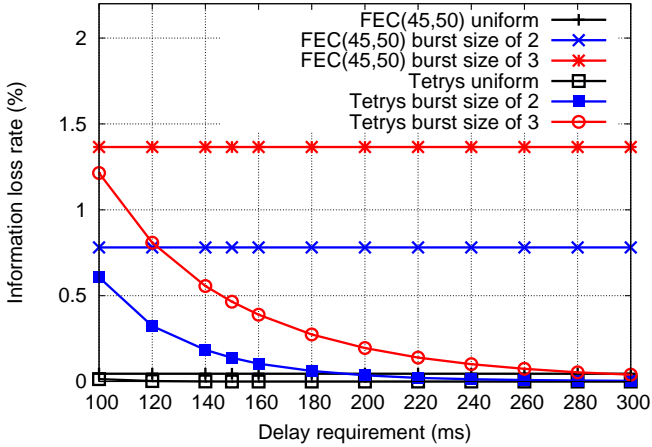}
\caption{Information loss rate of Tetrys and FEC(45,50) as a function of delay requirement with PLR of 3\% on both paths}
\label{fig:tetrys_delay_requirement}
\end{center}
\end{figure}

Fig. \ref{fig:tetrys_delay_requirement} shows the information loss rate of Tetrys and FEC(45,50) at PLR of 3\% on both paths and mean burst size of 3 packets. Since Tetrys is fully reliable, the lost packets are due to missed deadline. Thus, the information loss rate of Tetrys is reduced with the relaxation of the delay requirement. This implies that the gain of Tetrys against FEC is increased if the delay constraint is relaxed. It is noted that we use EMS as load splitting scheme to demonstrate the better performance of Tetrys against FEC, we believe that Tetrys still outperform FEC in any load splitting scheme.

\subsection{Propagation delay differences}
\label{sec:propa_delay_diff}
In \cite{EMS09}, the authors did not perform the tests where the available paths have different propagation delays. We compare Tetrys and FEC in case of 3 paths with different settings to \ref{sec:compa_fec_tetrys}. The PLR on each path is 14\%, 10\% and 12\%, respectively. The redundancy ratio is set to 25\% which is equivalent to FEC(15,20), FEC(24,32), FEC(30,40) and FEC(45,60). The simulation duration is 1000s with OLS Adapt Window of 1 second. Kurant showed in \cite{Kurant11} that the propagation delay differences between paths reach several tens of milliseconds. Thus, we vary the propagation delay on each path from 50 to 80ms so that no path has the same delay to the others and the maximum delay difference between paths is 30ms. This results in 24 simulations.

\begin{figure}[!htb]
\begin{center}
\includegraphics[scale=0.70]{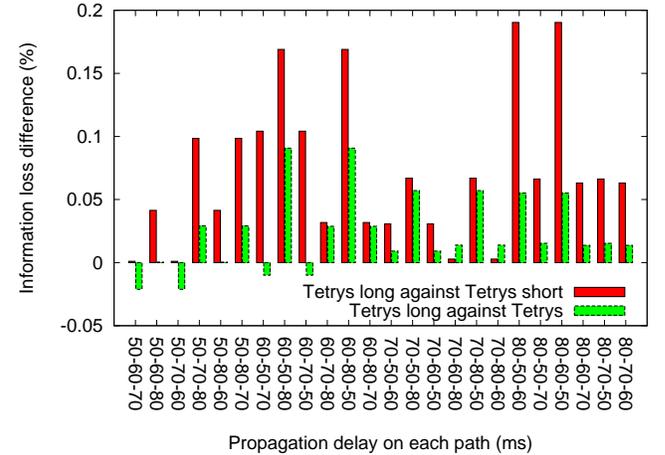}
\caption{``Tetrys long'' performance against ``Tetrys short'' and ``Tetrys``}
\label{fig:tetrys_compa_gb2}
\end{center}
\end{figure}

First, we compare different strategies of sending Tetrys repair packet (see Table \ref{tab:tetrys_strategies}). Fig. \ref{fig:tetrys_compa_gb2} shows the difference in information loss rate of ``Tetrys long'' against ``Tetrys short'' and ``Tetrys'' strategies for the mean burst size of 2 packets. The positive value means that the information loss rate of ``Tetrys long'' is less than the compared strategy (``Tetrys long'' is better) and vice versa. It is clear that ``Tetrys long'' strategy outperforms other strategies in most cases. Table \ref{tab:tetrys_strategies_result} shows the mean information loss rate and standard deviation of 24 simulations in case of uniform and mean burst size of 2 and 3 packets. ``Tetrys long'' strategy shows better results in all cases. At uniform losses and mean burst size of 2, 3 packets, ``Tetrys long'' gains 50\%, 24\% and 6\%, respectively, against the best strategy among ``Tetrys short'' and ``Tetrys''. These results confirm our analyis in \ref{sec:tetrys_decoupling}. Thus, we consider Tetrys as ``Tetrys long'' strategy from now on.

\begin{table}[h!]
\caption{Mean information loss rate and standard deviation of different strategies of sending Tetrys repair packets at uniform losses and burst losses with mean size of 2 and 3 packets}
\begin{center}
	\begin{tabular}{|l|l|l|l|}
	\hline
		 & Uniform & Burst size of 2 & Burst size of 3 \\ \hline

		Tetrys long & 0.0004\% $\pm$ 0.00056 &  0.083\% $\pm$ 0.021 & 0.47\% $\pm$ 0.15 \\ \hline
		Tetrys short &  0.015\% $\pm$ 0.039 & 0.15\% $\pm$ 0.06 & 0.5\% $\pm$ 0.1 \\ \hline
		Tetrys & 0.0008\% $\pm$ 0.0014 & 0.11\% $\pm$ 0.04 & 0.52\% $\pm$ 0.078 \\ \hline
	\end{tabular}
\end{center}
\label{tab:tetrys_strategies_result}
\end{table}

\begin{figure}[!htb]
\begin{center}
\includegraphics[scale=0.70]{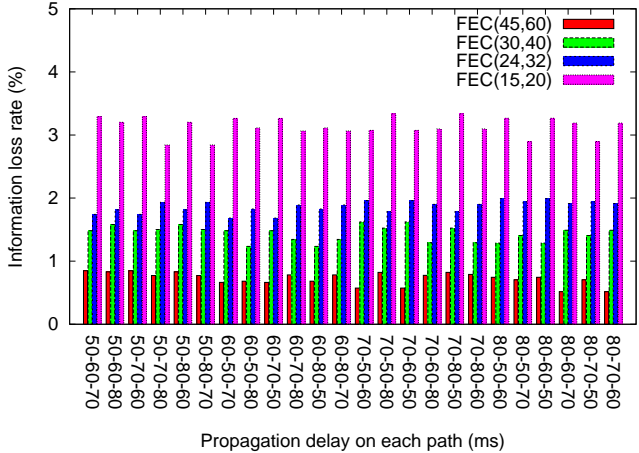}
\caption{Comparison between different FEC settings at mean burst size of 2 packets. PLR on each path is 14\%, 10\%, 12\%. Redundancy ratio is 25\%}
\label{fig:fec_compa_gb2}
\end{center}
\end{figure}

\begin{figure}[!htb]
\begin{center}
\includegraphics[scale=0.70]{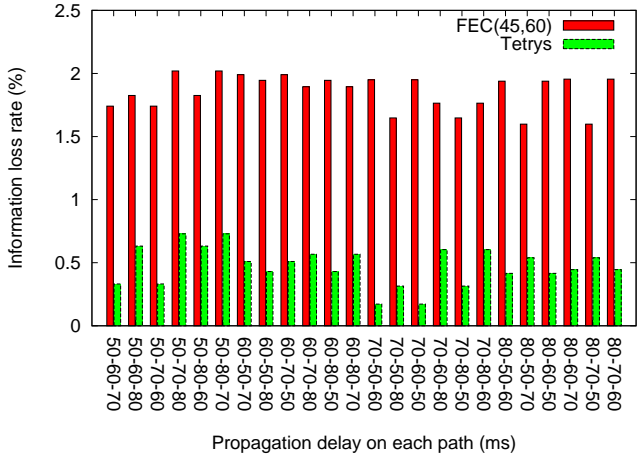}
\caption{FEC(45,60) vs Tetrys at mean burst size of 3 packets. PLR on each path is 14\%, 10\%, 12\%. Redundancy ratio is 25\%}
\label{fig:fec45_60_vs_tetrys_long_gb3}
\end{center}
\end{figure}

We then compare Tetrys with different FEC settings (FEC(15,20), FEC(24,32), FEC(30,40), FEC(45,60)). Fig. \ref{fig:fec_compa_gb2} shows the information loss rate of different FEC settings. The larger FEC block size makes FEC more tolerant to burst losses but leads to more delay to recover the lost packets. Fig. \ref{fig:fec45_60_vs_tetrys_long_gb3} shows the comparison between Tetrys and FEC(45,60), the best FEC among 4 settings, at mean burst size of 3 packets. Tetrys outperforms FEC(45,60) regardless the propagation delay on each path.

Table \ref{tab:fec_settings_result} shows the results of different FEC settings and Tetrys at both uniform losses and burst losses with mean size of 2 and 3 packets. We can see that Tetrys has a significant gain in information loss rate compared to FEC. Specifically, Tetrys has an average gain of 75\% in information loss rate against the best FEC at mean burst size of 3 packets.

\begin{table}[h!]
\caption{Mean and standard deviation information loss rate with different FEC settings and Tetrys}
\begin{center}
	\begin{tabular}{|l|l|l|l|}
	\hline
		 & Uniform & Burst size of 2 & Burst size of 3 \\ \hline
		FEC(15,20) & 0.53\% $\pm$ 0.14 &  3.14\% $\pm$ 0.15 & 4.77\% $\pm$ 0.22 \\ \hline
		FEC(24,32) & 0.18\% $\pm$ 0.051 & 1.87\% $\pm$ 0.09 & 3.55\% $\pm$ 0.18 \\ \hline
		FEC(30,40) & 0.11\% $\pm$ 0.041 & 1.44\% $\pm$ 0.12 & 2.81\% $\pm$ 0.19 \\ \hline
		FEC(45,60) & 0.028\% $\pm$ 0.017 & 0.73\% $\pm$ 0.099 & 1.86\% $\pm$ 0.13 \\ \hline
		Tetrys & 0.0004\% $\pm$ 0.00056 &  0.083\% $\pm$ 0.021 & 0.47\% $\pm$ 0.15 \\ \hline
	\end{tabular}
\end{center}
\label{tab:fec_settings_result}
\end{table}

\section{EMS algorithm improvement}
\label{sec:ems_algorithm_improvement}
The original OLS algorithm (see pseudo code \ref{alg:ols_original}) shows very good results. However, it does not adapt well to the network dynamics. In fact, assuming that the OLS is increasing the load on path 1, the loss rate on path 2 reduces significantly and is lower than path 1. This might lead to the better information loss rate, the original OLS scheme is still in a repeat-until loop and continues to increase the load on path 1. This make the OLS scheme goes farther from the new optimal load splitting, while it is better to stop increasing the load on path 1 and to increase the load on path 2. Thus, we propose to add a pre-defined threshold of loss rate $\theta$. At each period, the scheme compares the loss rate on each path with the one in previous period, if the absolute difference is greater than $\theta$, the OLS scheme quits the repeat-until loop and re-sorts the paths. The improved OLS scheme is depicted in the pseudo code \ref{alg:ols_modified}. 

\begin{algorithm}
\caption{Modified OLS}
\label{alg:ols_modified}
\begin{algorithmic}[1]
\STATE Compute the asymptotic optimal solution and split the load accordingly 
\STATE Sort the paths in the increasing order of loss rate
\REPEAT 
\STATE Pick the first path in the list

\REPEAT 
\IF {\textit{the absolute difference of loss rate on one path exceeds the pre-defined threshold $\theta$}}
\STATE \textbf{goto} Step 2
\ENDIF
\STATE Increase the load on the chosen path by pre-defined $\Delta_L$
\STATE Decrease the load on each of remaining paths by a fraction of $\delta$, proportional to their respective loss rates
\UNTIL{measured information loss rate increases}
\STATE Remove the chosen path from the list
\STATE Revert to the previous load splitting
\UNTIL{the path list is empty}
\STATE \textbf{goto} Step 2
\end{algorithmic}
\end{algorithm}

\begin{figure}[!htb]
\begin{center}
\includegraphics[scale=0.70]{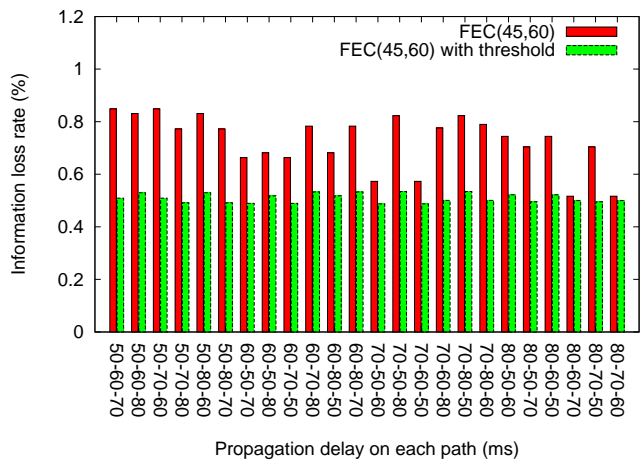}
\caption{FEC(45,60) without threshold and with threshold $\theta$ = 5\% at mean burst size of 2 packets}
\label{fig:fec45_60_threshold5_gb2}
\end{center}
\end{figure}

\begin{figure}[!htb]
\begin{center}
\includegraphics[scale=0.70]{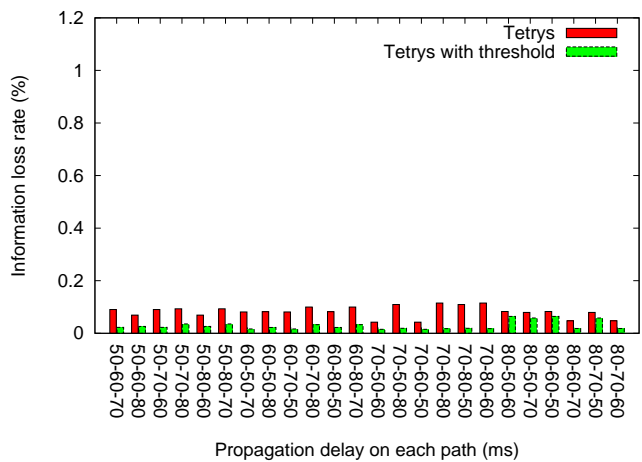}
\caption{Tetrys without threshold and with threshold $\theta$ = 5\% at mean burst size of 2 packets}
\label{fig:tetrys_long_threshold5_gb2}
\end{center}
\end{figure}

With the same settings as in \ref{sec:propa_delay_diff}, we re-run the simulations with a pre-defined threshold $\theta$ = 5\%. The information loss rate of both FEC(45,60) and Tetrys with threshold is lower than the one without threshold in case of mean burst size of 2 packets (Fig. \ref{fig:fec45_60_threshold5_gb2} and \ref{fig:tetrys_long_threshold5_gb2}). Table \ref{tab:threshold} shows the improvement in information loss rate with modified EMS scheme in both uniform and burst losses. At mean burst size of 2 packets, FEC(45,60) and Tetrys with threshold has an average gain of 30\% and 65\%, respectively, compared to the case without threshold. While FEC(45,60) and Tetrys with modified EMS scheme have an average gain of 21\% and 49\%, respectively in comparison to the original EMS scheme at mean burst size of 3 packets. These simulations show that Tetrys achieves much lower information loss rate with modified OLS algorithm although it has a very small information loss rate using the original one.


\begin{table}[h!]
\caption{Mean and standard deviation information loss rate with and without threshold}
\begin{center}
	\begin{tabular}{|c|l|l|l|}
	\hline
		\multicolumn{2}{|c|}{} & Without threshold & Threshold $\theta$ = 5\% \\ \hline
		\multirow{5}{*}{Uniform} & FEC(15,20) & 0.53\% $\pm$ 0.14 & 0.4\% $\pm$ 0.01 \\ \cline{2-4}
		 & FEC(24,32) & 0.18\% $\pm$ 0.051 & 0.12\% $\pm$ 0.0044\\ \cline{2-4}
		 & FEC(30,40) & 0.11\% $\pm$ 0.041 & 0.056\% $\pm$ 0.0053\\ \cline{2-4}
		 & FEC(45,60) & 0.028\% $\pm$ 0.017 & 0.014\% $\pm$ 0.0058\\ \cline{2-4}
		 & Tetrys & 0.0004\% $\pm$ 0.00056 & 0.00016\% $\pm$ 0.00025\\ \hline
		 & FEC(15,20) & 3.14\% $\pm$ 0.15 & 2.71\% $\pm$ 0.019 \\ \cline{2-4}
		Burst & FEC(24,32) & 1.87\% $\pm$ 0.09 & 1.58\% $\pm$ 0.02 \\ \cline{2-4}
		size & FEC(30,40) & 1.44\% $\pm$ 0.12 & 1.12\% $\pm$ 0.014 \\ \cline{2-4}
		of 2 & FEC(45,60) & 0.73\% $\pm$ 0.099 & 0.51\% $\pm$ 0.017 \\ \cline{2-4}
		 & Tetrys & 0.083\% $\pm$ 0.021 & 0.029\% $\pm$ 0.016 \\ \hline
		 & FEC(15,20) & 4.77\% $\pm$ 0.22 & 4.45\% $\pm$ 0.024 \\ \cline{2-4}
		Burst & FEC(24,32) & 3.55\% $\pm$ 0.18 & 3.07\% $\pm$ 0.022 \\ \cline{2-4}
		size & FEC(30,40) & 2.81\% $\pm$ 0.19 & 2.44\% $\pm$ 0.038 \\ \cline{2-4}
		of 3 & FEC(45,60) & 1.86\% $\pm$ 0.13 & 1.47\% $\pm$ 0.036 \\ \cline{2-4}
		 & Tetrys & 0.47\% $\pm$ 0.15 & 0.24\% $\pm$ 0.071 \\ \hline
	\end{tabular}
\end{center}
\label{tab:threshold}
\end{table}

\section{Conclusions and future work}
\label{sec:conclusion}

In this paper, we introduced an on-the-fly coding scheme named Tetrys to real-time multipath streaming. With the same load splitting scheme, the EMS scheme presented in \cite{EMS09}, we have shown that Tetrys consistently has significant reduction in information loss rate compared to the FEC approach in both uniform and burst losses. We showed that the decoupling between load allocation and Tetrys redundancy traffic improves the performance in terms of loss rate after packet recovery. The Tetrys repair packets are preferably sent to the path with longer propagation delay shows best performance. Furthermore, we showed that the EMS scheme can be improved to provide better results. By introducing a threshold parameter, the modified EMS scheme adapts well to the network dynamics and showed a significant reduction in information loss rate compared to the original one. For future work, we plan to analyze the multipath streaming in more realistic contexts (e.g., 2 paths with Wi-Fi and 3G/LTE) and to validate the results with video data.

\section{Acknowledgments}
This work was supported by the French ANR grant ANR-VERS-019-02 (ARSSO project).

\bibliographystyle{IEEEtran}
\bibliography{biblio.bib}

\begin{thebibliography}{10}
\providecommand{\url}[1]{#1}
\csname url@samestyle\endcsname
\providecommand{\newblock}{\relax}
\providecommand{\bibinfo}[2]{#2}
\providecommand{\BIBentrySTDinterwordspacing}{\spaceskip=0pt\relax}
\providecommand{\BIBentryALTinterwordstretchfactor}{4}
\providecommand{\BIBentryALTinterwordspacing}{\spaceskip=\fontdimen2\font plus
\BIBentryALTinterwordstretchfactor\fontdimen3\font minus
  \fontdimen4\font\relax}
\providecommand{\BIBforeignlanguage}[2]{{%
\expandafter\ifx\csname l@#1\endcsname\relax
\typeout{** WARNING: IEEEtran.bst: No hyphenation pattern has been}%
\typeout{** loaded for the language `#1'. Using the pattern for}%
\typeout{** the default language instead.}%
\else
\language=\csname l@#1\endcsname
\fi
#2}}
\providecommand{\BIBdecl}{\relax}
\BIBdecl

\bibitem{bestpath_vs_multipath}
D.~G. Andersen, A.~C. Snoeren, and H.~Balakrishnan, ``Best-path vs. multi-path
  overlay routing,'' in \emph{IN PROC. ACM SIGCOMM Internet Measurement
  Conference}, 2003, pp. 91--100.

\bibitem{Golubchik02multipath}
L.~Golubchik, J.~C. Lui, T.~F. Tung, A.~L. Chow, W.~j.~Lee, G.~Franceschinis,
  and C.~Anglano, ``Multi-path continuous media streaming: What are the
  benefits?'' 2002.

\bibitem{Fashandi}
S.~Fashandi, S.~O. Gharan, and A.~K. Khandani, ``Path diversity over packet
  switched networks: performance analysis and rate allocation,'' \emph{IEEE/ACM
  Trans. Netw.}, vol.~18, pp. 1373--1386, October 2010.

\bibitem{Jurca_mediaflow}
D.~Jurca and P.~Frossard, ``Media flow rate allocation in multipath networks,''
  \emph{IEEE Transactions on Multimedia}, p. 12271240, 2007.

\bibitem{Jurca_forwarderror}
D.~Jurca, P.~Frossard, and A.~Jovanovic, ``Forward error correction for
  multipath media streaming,'' \emph{IEEE Trans. Circuits and Systems for Video
  Technology}, pp. 1315--1326, 2009.

\bibitem{EMS09}
A.~L.~H. Chow, H.~Yang, C.~H. Xia, M.~Kim, Z.~Liu, and H.~Lei, ``{EMS: Encoded
  Multipath Streaming for Real-time Live Streaming Applications},'' in
  \emph{IEEE International Conference on Network Protocols, Princeton, NJ, USA,
  13-16 October 2009}.\hskip 1em plus 0.5em minus 0.4em\relax IEEE Computer
  Society, 2009, pp. 233--243.

\bibitem{Nguyen03pathdiversity}
T.~Nguyen and A.~Zakhor, ``{Path Diversity with Forward Error Correction (PDF)
  System for Packet Switched Networks},'' in \emph{in Proceedings of IEEE
  INFOCOM}, 2003, pp. 663--672.

\bibitem{arq_nc}
J.~Kumar~Sundararajan, D.~Shah, and M.~Medard, ``{ARQ for network coding},'' in
  \emph{IEEE International Symposium on Information Theory}, july 2008, pp.
  1651 --1655.

\bibitem{Lacan}
J.~Lacan and E.~Lochin, ``{Rethinking reliability for long-delay networks},''
  in \emph{Proceedings of International Workshop on Satellite and Space
  Communications IWSSC}, Oct. 2008.

\bibitem{Tetrys}
P.-U. Tournoux, E.~Lochin, J.~Lacan, A.~Bouabdallah, and V.~Roca, ``On-the-fly
  coding for time-constrained applications,'' \emph{IEEE Transactions on
  Multimedia}, vol.~13, 2011.

\bibitem{RFS}
G.~Cheung, ``{Near-optimal multipath streaming of H.264 using reference frame
  selection},'' in \emph{Proceedings of International Conference on Image
  Processing}, sept. 2003.

\bibitem{Kurant11}
M.~Kurant, ``Exploiting the path propagation time differences in multipath
  transmission with fec,'' \emph{IEEE Journal on Selected Areas in
  Communications}, vol.~29, no.~5, pp. 1021--1031, 2011.

\bibitem{ns}
``The {N}etwork {S}imulator http://www.isi.edu/nsnam/ns/index.html.''

\bibitem{itu}
``{ITU-T recommendation G.114},'' International Telecommunication Union, Tech.
  Rep., 2009.

\bibitem{Wenger03}
S.~Wenger, ``{H.264/AVC over IP},'' \emph{IEEE Transactions on Circuits and
  Systems for Video Technology}, vol.~13, pp. 645--656, 2003.

\end{thebibliography}

\end{document}